\documentclass[twocolumn,showpacs,preprintnumbers,amsmath,amssymb,superscriptaddress]{revtex4}
\usepackage{graphicx}
\usepackage{dcolumn}
\usepackage{bm}
\usepackage[chatter]{rotating}

\def\bea{\begin{eqnarray}}
\def\eea{\end{eqnarray}}

\begin{document}


\title{Azimuthal quadrupole correlation from gluon interference in 200 GeV and 7 TeV p+p collisions}

\affiliation{Department of Physics, The University of Texas at Austin, Austin, Texas 78712 USA}

\author{R. L. Ray}\affiliation{Department of Physics, The University of Texas at Austin, Austin, Texas 78712 USA}

\date{\today}

\begin{abstract}

The Balitskii-Fadin-Kuraev-Lipatov (BFKL) multi-Pomeron model of Levin and Rezaeian, with extension to the gluon saturation region, is applied to long-range pseudorapidity correlations on relative azimuth for low momentum final-state hadrons produced in $\sqrt{s}$ = 200~GeV and 7~TeV p+p collisions. The multi-Pomeron exchange probabilities in the model were estimated by fitting the minimum-bias p+p multiplicity frequency distributions. The multi-Pomeron model prediction for the amplitude of the minimum-bias average quadrupole correlation, proportional to $\cos 2(\phi_1 - \phi_2)$, is consistent with the 200~GeV data when theoretically expected gluon saturation momentum scales are used. Correlation predictions for the high multiplicity 7~TeV p+p collision data are also consistent with the long-range pseudorapidity correlations at small relative azimuth observed in the data. The results presented here show that the present application of a multiple parton-shower, gluon interference mechanism for generating the long-range pseudorapidity, azimuthal quadrupole correlation is not excluded by the data.

\end{abstract}


\maketitle

\section{Introduction}
\label{SecI}

The observation of long-range two-particle angular correlations on relative
pseudorapidity ($\eta$) in high multiplicity p+p collisions at $\sqrt{s}$ = 7~TeV
by the CMS collaboration~\cite{CMSpp7TeV} motivated considerable interest
in searching for similar long-range correlations in other, ``elementary''
hadronic collision systems. These systems include p+Pb at the Large Hadron Collider (LHC) and p+p and d+Au at the Relativistic Heavy Ion Collider (RHIC).
Three independent, phenomenological analyses~\cite{TomQuadpp7TeV,Bozek,LRQuadpp7TeV}
of the CMS correlation data from p+p collisions
showed that the correlation structures, and in particular the long-range $\eta$
feature, were well described by models which included a second-order cosine-series
element $\cos{2(\phi_1 - \phi_2)}$, where $\phi_{1,2}$ are the azimuthal angles of two, arbitrary final-state hadrons. This function is an azimuthal quadrupole in cylindrical
coordinates. Recent measurements from the LHC experiments~\cite{ATLASpPb,ALICEpPb,CMSpPb}
found evidence of similar, long-range $\eta$ correlations in the $\sqrt{s_{NN}}$ = 5.02~TeV p+Pb collision
data which increased in amplitude with increasing multiplicity. Analysis by the ALICE
collaboration~\cite{ALICEpPb} showed that this multiplicity dependent correlation
structure could be accurately described with an azimuthal quadrupole.  The PHENIX collaboration~\cite{PHENIXdAu}
reported similar quadrupole structure in the angular correlation data for
$\sqrt{s_{NN}}$ = 200~GeV d+Au minimum-bias collisions. Analysis of preliminary
multiplicity dependent correlations from $\sqrt{s}$ = 200~GeV p+p collisions (STAR collaboration~\cite{TomDuncan})
also obtained statistically significant evidence of a quadrupole structure which increased
quadratically with multiplicity.  Long-range, $\eta$-independent quadrupole correlations
in heavy-ion collisions have been observed for decades~\cite{AAv2}. It now appears that such correlations
are ubiquitous in the intermediate momentum, final-state charged hadron yields from high-energy hadronic collisions.

The CMS p+p correlation data also motivated theoretical interest in perturbative
quantum chromo-dynamics (pQCD) approaches which might explain these long-range correlations.
Calculations based on a color-dipole model~\cite{Boris}, gluon interference from connected,
multi-Pomeron exchange diagrams~\cite{LR}, and gluon interference
from multiple radiating {\em glasma} flux tubes~\cite{glasma,RajuDusling} all predicted a significant
quadrupole correlation in high energy p+p and p+A (nucleus) collisions. Quantitative descriptions of the long-range $\eta$ correlations on relative azimuth observed in the 7~TeV p+p and the 5.02~TeV p+Pb collision data from the CMS collaboration based on the glasma model were reported in~\cite{glasma,RajuDusling}.
Explanations based on collective flow generated by secondary collisions between partons produced in the initial-stage of p+p and p+A collisions have also been proposed~\cite{Bozek,pPbFlow}.

In Ref.~\cite{LR} Levin and Rezaeian showed that the BFKL Pomeron description of p+p collisions in leading $\log(s)$ approximation, when extended into the gluon saturation kinematic region, produces long-range correlations on relative $\eta$ which display an azimuthal quadrupole structure, $\cos2(\phi_1 - \phi_2)$, whenever two or more Pomeron exchanges occur in individual collision {\em events}. The produced single-gluon distribution includes a $\cos2\phi$ dependence relative to the direction of the transverse momentum exchange vector $\vec{Q}_{\rm T}$. The two-gluon azimuthal correlations arise from quantum interference between the soft gluons radiated from two or more parton showers which couple different pairs of initial-state partons in the colliding protons. Local parton hadron duality (LPHD)~\cite{LPHD} can be assumed to relate the radiated, interfering gluons to the final-state hadrons.



The Levin and Rezaeian (LR) BFKL~\cite{BFKL} multi-Pomeron model (referred to as the LR-model in the following) contains several unspecified parameters and includes a number of approximations. The predicted correlations depend on the number of Pomeron exchanges, the multi-Pomeron exchange probabilities, and the gluon saturation momentum scale $Q_{\rm S}^2$, none of which is given in Ref.~\cite{LR}. With respect to approximations the radiated gluon densities were calculated in the small transverse momentum ($p_t$) limit and expanded in powers of $Q_{\rm T}$ to second-order. Pomeron-proton coupling was restricted to tree-level diagrams; triple Pomeron terms were omitted. Integrals over transverse momentum exchange in the gluon saturation region were assumed to be dominated by the saturation scale and were approximated as described in the following.

At both the RHIC and the LHC collision energies much of the inclusive particle production occurs for $p_t$ less than the expected saturation scale. As a result enhanced Pomeron exchange diagrams in the saturation region (those containing loops) contribute to the momentum integral $\langle q_{\rm T}^{-4} \rangle$~\cite{LR} over BFKL Pomeron densities, making precise calculations a challenge. Using Pomeron loop summation techniques~\cite{PomeronLoops} LR argued that these momentum integrals are dominated by $Q_{\rm S}^2$. Uncertainties were not provided however. Integration over transverse momentum $\vec{Q}_{\rm T}$ and the Pomeron-proton interaction vertex was carried out in the LR-model for the BFKL kinematic region. In the saturation region this integral was assumed to be dominated by $Q_{\rm S}^2$. The resulting range of values for quantity $\langle \langle Q_{\rm T}^4 \rangle \rangle $ in \cite{LR} provides a liberal estimate of its uncertainty.
 
In this work estimates of the Pomeron exchange probabilities and $Q_{\rm S}^2$, including uncertainties, are provided for 200~GeV p+p collision data from the STAR collaboration~\cite{TomDuncan} and the high multiplicity 7~TeV p+p collision data from the CMS collaboration~\cite{CMSpp7TeV}. Because of the restriction to low-$p_t$, the LR-model is only compared to the inclusive, $p_t$-integral correlation data. The primary goal of this analysis is to determine if the LR-model with its assumptions and uncertainties summarized above, plus the parameters obtained here, can be falsified by the quadrupole correlation measurements for p+p collisions at the RHIC and the LHC and, if so, is the present application of the LR-model falsified.

The capability of the multi-Pomeron exchange model to describe the multiplicity frequency distributions in high energy p+p collisions is demonstrated in Sec.~\ref{SecII}. Pomeron exchange probabilities for the LR-model are also estimated in Sec.~\ref{SecII}. The quadrupole correlation amplitude for an ensemble of p+p collisions with a varying number of multi-Pomeron exchanges is derived in Sec.~\ref{SecIII}. Saturation scale estimates are given in Sec.~\ref{SecIV}. Results for the 200~GeV and 7~TeV data are compared with experimental results in Secs.~\ref{SecV} and \ref{SecVI}, respectively, and are further discussed in Sec.~\ref{SecVII}. Conclusions are given in Sec.~\ref{SecVIII}.

\section{Pomeron exchange probabilities}
\label{SecII}

The Pomeron-exchange probabilities were estimated by first assuming that parton showers~\cite{LR} account for the entire soft particle production. Given this assumption the minimum-bias multiplicity frequency distribution for 200~GeV non-singly diffractive (NSD) p+p collisions~\cite{UA5} was fitted with an ensemble of collision events where each collision produces either 1, 2, 3 or more parton showers. In the LR-model each parton shower was assumed to produce a Poisson multiplicity frequency distribution of gluons with mean number equal to the minimum-bias average number of charged hadrons $\langle n_{\rm ch} \rangle / \Delta\eta = 2.5$ at $\sqrt{s}$ = 200~GeV~\cite{UA5,Tom-pp,neutrals} and $\langle n_{\rm ch} \rangle / \Delta\eta = 5.78$ at $\sqrt{s}$ = 7~TeV~\cite{CMSdNdeta} where $\Delta\eta$ is the pseudorapidity acceptance. At these energies a non-negligible fraction of the multiplicity is produced by semi-hard scattering which is not included in the parton showers. To account for semi-hard scattering the two-component multiplicity model in Ref.~\cite{KN} and an analysis of the multiplicity-dependent transverse momentum ($p_t$) spectrum data for 200~GeV p+p collisions~\cite{Tom-pp} were used.

\begin{figure*}[t]
\includegraphics[keepaspectratio,width=4.0in]{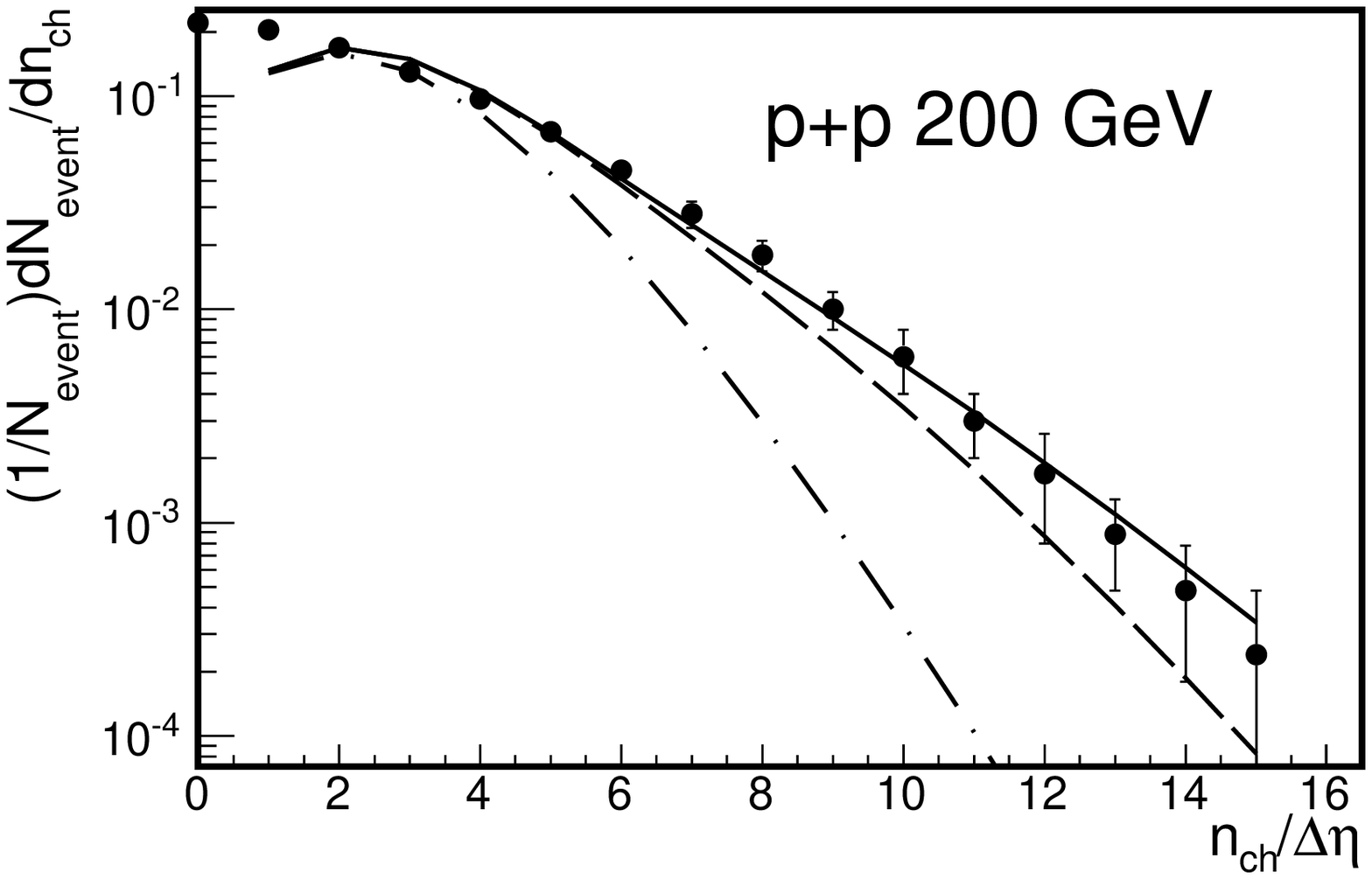}
\put(-225,40){\bf (a)}
\includegraphics[keepaspectratio,width=2.8in]{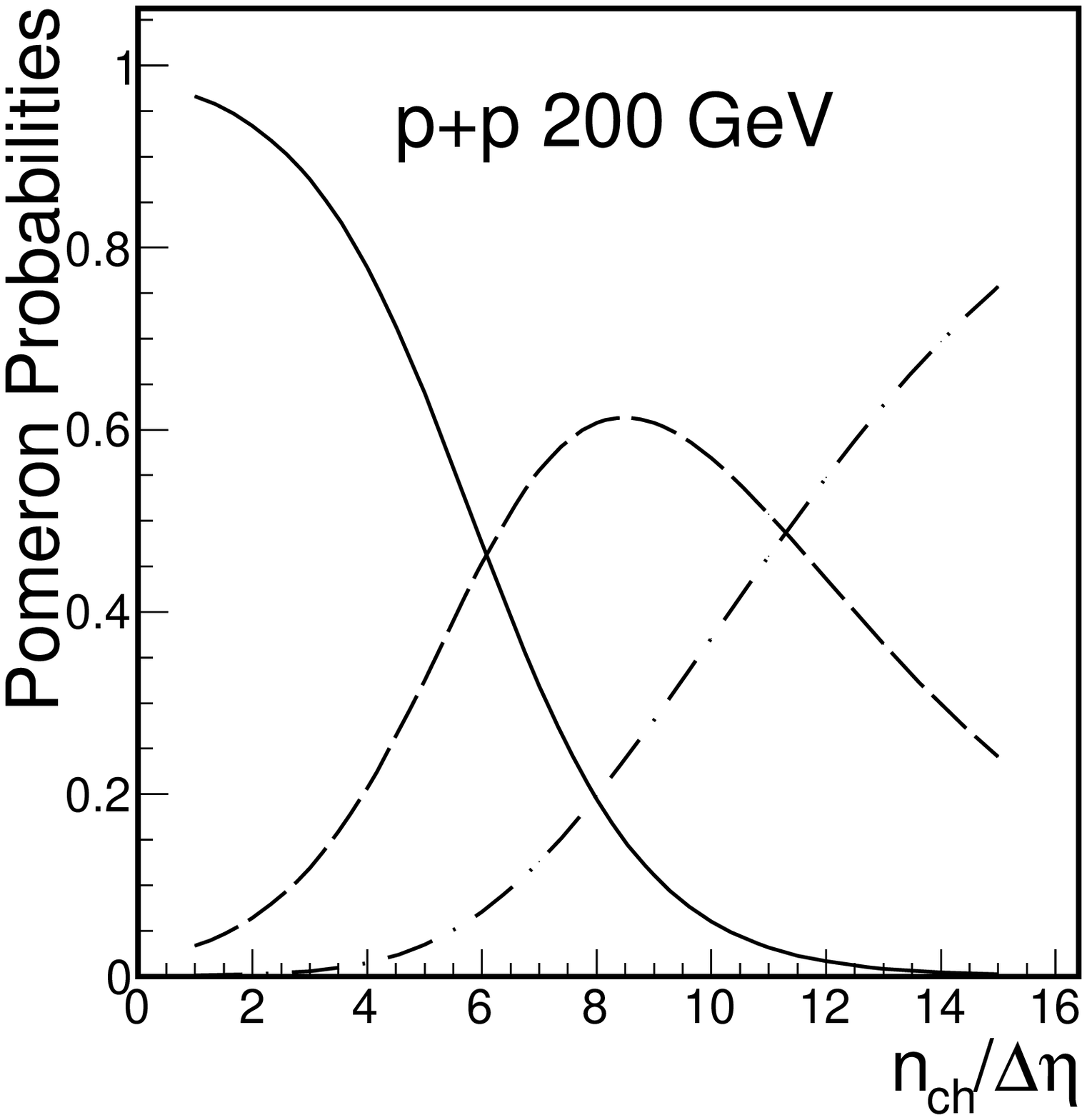}
\put(-155,50){\bf (b)}
\caption{\label{Fig1}
Panel (a): Minimum-bias multiplicity frequency distribution data for $\sqrt{s}$ = 200 GeV p+p NSD data for $|\eta| \leq 0.5$ from UA5~\cite{UA5} and Pomeron exchange model fits (curves) described in the text. Panel (b): Corresponding multi-Pomeron exchange probabilities for collisions as functions of multiplicity $n_{\rm ch}$. Probabilities for $m = 1,2,3$ are shown by the solid, dashed and dashed-dotted curves, respectively.}
\end{figure*}

\begin{figure}[h]
\includegraphics[keepaspectratio,width=3.7in]{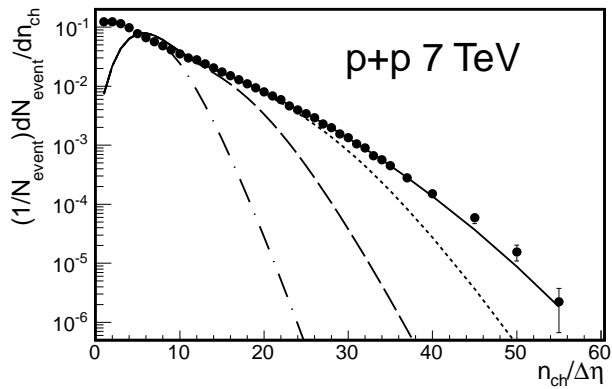}
\caption{\label{Fig2}
Same as Fig.~\ref{Fig1}a except for 7~TeV p+p collisions from CMS~\cite{CMSFreq}. Pomeron exchange model fits (curves) are described in the text.}
\end{figure}

In the 200~GeV $p_t$ spectrum analysis of Ref.~\cite{Tom-pp} the charged particle multiplicity was expressed as the sum $n_{\rm ch} = n_{\rm s} + n_{\rm h}$, where $n_{\rm s}$ and $n_{\rm h}$ are the soft (Pomeron shower) and hard (semi-hard minimum-bias jet) particle production multiplicities, respectively, within the $\eta$ acceptance for 2$\pi$ azimuth and $p_t > 0.15$~GeV/$c$. For each sub-set of collisions having fixed $n_{\rm ch}$, the ratio of the mean hard-component multiplicity to the mean soft-component multiplicity was found to scale linearly with $n_{\rm ch}$ where~\cite{Tom-pp}
\bea
\frac{\bar{n}_{\rm h} (n_{\rm ch})}{\bar{n}_{\rm s} (n_{\rm ch})}
 & = &
\frac{\alpha n_{\rm ch} }{\epsilon \Delta\eta}
\label{Eq1}
\eea
and $\alpha = 0.0105 \pm 0.0005$, $\epsilon$ is the inverse tracking efficiency at mid-rapidity ($2.0 \pm 0.02$), pseudorapidity acceptance $\Delta\eta = 1.0$ corresponding to tracking acceptance $|\eta| \leq 0.5$, and $n_{\rm ch}$ is corrected for efficiency and acceptance. Soft and hard-component mean values are given by
\bea
\bar{n}_{\rm s} (n_{\rm ch}) & = & n_{\rm ch}/ \left( 1 +
\alpha n_{\rm ch}/(\epsilon \Delta\eta) \right), \nonumber \\
\bar{n}_{\rm h}(\bar{n}_{\rm s}) & = & \frac{\alpha \bar{n}_{\rm s}^2
/(\epsilon \Delta\eta) } {1 - \alpha \bar{n}_{\rm s}/(\epsilon \Delta\eta) }.
\label{Eq2}
\eea

The event-wise multiplicity frequency distribution on $n_{\rm s}$ in the LR-model was expressed as a sum of Poisson distributions, ${\cal P}(n_{\rm s},\bar{n}_{\rm s})$ with mean $\bar{n}_{\rm s} = m \langle n_{\rm ch} \rangle$ where $m = 1,2,3 \cdots$. The hard-component frequency distribution, ${\cal P}(n_{\rm h},\bar{n}_{\rm h})$, was also assumed to be Poisson. The frequency distribution on total charged particle multiplicity was obtained by projecting the joint probability distribution on $(n_{\rm s},n_{\rm h})$ onto $n_{\rm ch}$. Using unit-normalized distributions and probabilities the LR Pomeron shower plus two-component model for the multiplicity frequency distribution is given by
\bea
\frac{1}{N_{\rm events}} \frac{dN_{\rm events}} {dn_{\rm ch}} 
 \nonumber \\
 & & \hspace{-1.0in} =
\sum_{n_{\rm s}} {\cal P}[n_{\rm ch} - n_{\rm s}, \bar{n}_{\rm h}(n_{\rm s})]
\sum_{m=1}^{M} P_m {\cal P}(n_{\rm s},m \langle n_{\rm ch} \rangle ),
\label{Eq3}
\eea
where $\bar{n}_{\rm h}(n_{\rm s})$ is obtained from Eq.~(\ref{Eq2}) and $P_m$ is the probability that a p+p collision in the minimum-bias event ensemble includes $m$ parton showers where $\sum_m P_m = 1$. Probabilities $P_m$ were adjusted to fit the UA5 measurements for $|\eta| \leq 0.5$ and 1.5 for the 200~GeV data~\cite{UA5}. 

The UA5 data and best fit are shown in Fig.~\ref{Fig1}a where the dashed-dotted, dashed and solid curves represent, respectively, the contributions from collisions with one Pomeron exchange only, with one and two Pomeron exchanges, and with the combined one, two and three Pomeron exchanges. Higher-order exchange processes ($m > 3$) were not required to describe these data within the measured multiplicity range. The model does not describe the distribution at small $n_{\rm ch} < \langle n_{\rm ch} \rangle$. The fitting was therefore optimized within the larger $n_{\rm ch} \geq \langle n_{\rm ch} \rangle$ domain where multi-Pomeron exchanges, which determine the quadrupole correlation, are more probable. Each model distribution in Fig.~\ref{Fig1}a includes hard scattering contributions which are most significant at higher multiplicities~\cite{Tom-pp}. The estimated probabilities are $P_1 = 0.802$, $P_2 = 0.168$ and $P_3 = 0.030$ for the $|\eta| \leq 0.5$ data. A similar analysis was applied to the UA5 $|\eta| \leq 1.5$ frequency distribution with corresponding probabilities 0.749, 0.218 and 0.033. The probabilities for collision events having one, two and three-Pomeron exchange as a function of $n_{\rm ch}$ for the $|\eta| \leq 0.5$ fit results are shown in Fig.~\ref{Fig1}b by the solid, dashed and dashed-dotted curves, respectively.

The multiplicity frequency distribution data for 7~TeV p+p NSD collisions for particles produced with $|\eta| < 0.5$ measured by the CMS collaboration~\cite{CMSFreq} were similarly described by the above model. However, the hard- to soft-component multiplicity ratio $\alpha$ has not been determined at this collision energy as was done in Ref.~\cite{Tom-pp} for the 200~GeV data. The same value $\alpha$ = 0.0105 was assumed. However, this ratio may change at higher collision energies depending on the evolution of the gluon distribution~\cite{HERA} and the soft-particle production mechanism. An increase in $\alpha$ by a factor of 2 to 3 was permitted ($\alpha$ = 0.02 and 0.03) based on the evolution of the gluon distribution with $Q^2$ at relevant values of $x$ for 200~GeV and 7~TeV collisions (10$^{-2}$ and 10$^{-4}$, see Sec.~\ref{SecIV}). Comparable quality fits to the frequency distribution data were obtained for each value of $\alpha$, however the three- and higher-order Pomeron exchange probabilities were strongly affected by the choice of $\alpha$. The CMS data and best fit with $\alpha$ = 0.0105 are shown in Fig.~\ref{Fig2}. With this choice of $\alpha$ the best model fit to the data required collisions with up to four Pomeron exchanges. The contributions from collisions with only one Pomeron exchange, with one and two Pomeron exchanges, with one, two and three Pomeron exchanges, and with as many as four Pomeron exchanges are shown by the dashed-dotted, dashed, dotted and solid lines, respectively, in Fig.~\ref{Fig2}. The differing Pomeron exchange probabilities resulting from the choices for $\alpha$ are included in the uncertainties of the LR-model correlation predictions discussed in Sec.~\ref{SecVI}.

Figs.~\ref{Fig1} and \ref{Fig2} demonstrate the capability of the Pomeron shower model to accurately describe the multiplicity frequency distributions, a necessary requirement for the model. The resulting multi-Pomeron exchange probabilities as a function of multiplicity are used in the following sections.

\section{Quadrupole correlation}
\label{SecIII}

The one- and two-gluon densities in Eqs.~(22)-(26) of Ref.~\cite{LR}, the scaling of the soft-component of the single-gluon emission with the number of parton showers, and the scaling of the number of correlated gluon pairs with the number of pairs of parton showers per collision [Eq.~(27) in \cite{LR}] form the basis for the present calculations. The one- and two-gluon densities produced by $k$ parton showers in p+p collisions which have a total of $m$ parton showers, where $k \leq m$, are denoted by $\rho_{k(mI\!\!P)}(n_{\rm s};1)$ and $\rho_{k(mI\!\!P)}(n_{\rm s};1,2)$, respectively. In these definitions $n_{\rm s}$ is the soft-component of the charged particle multiplicity (assuming LPDH)~\cite{LPHD,Tom-pp,KN,neutrals} in acceptance $\Delta\eta$, and labels 1 and 2 represent the coordinates of the radiated gluons.

The soft-component of the one-gluon density can be expressed, with the above assumptions~\cite{fluctuations}, as
\bea
\rho_{k(mI\!\!P)}(n_{\rm s};1) & = & k \rho_{1(mI\!\!P)}(n_{\rm s};1) \approx
n_{\rm s} \frac{k}{m} \hat{\rho}_{\rm s}(1),
\label{Eq4}
\eea
where $\hat{\rho}_{\rm s}(1)$ is the unit-normal, soft-component single-gluon density, assumed equal (in LPHD) to the single charged particle density~\cite{Tom-pp}. For collisions with two parton showers and soft-component multiplicity $n_{\rm s}^{\prime}$ the two-gluon density in the LR-model is
\bea
\label{Eq5}
\rho_{2(2I\!\!P)}(n_{\rm s}^{\prime};1,2) & = &
\frac{n_{\rm s}^{\prime} - 1}{n_{\rm s}^{\prime}} \rho_{2(2I\!\!P)}(n_{\rm s}^{\prime};1) \rho_{2(2I\!\!P)}(n_{\rm s}^{\prime};2) \nonumber \\
 & + & \Delta\rho_{2(2I\!\!P)}(n_{\rm s}^{\prime};1,2) \\
\label{Eq6}
\rho_{2(2I\!\!P)}(n_{\rm s}^{\prime};1,2)
 & = & \frac{n_{\rm s}^{\prime} - 1}{n_{\rm s}^{\prime}}
\rho_{2(2I\!\!P)}(n_{\rm s}^{\prime};1) \rho_{2(2I\!\!P)}(n_{\rm s}^{\prime};2) \nonumber \\
 & \times & \left[ 1 + f_{2(2I\!\!P)}(Q^2_{\rm S};1,2) \right], \\ 
\label{Eq7}
f_{2(2I\!\!P)}(Q^2_{\rm S};1,2) \nonumber \\
 & & \hspace{-1.0in} \approx  \frac{1}{2} p^2_{t_1} p^2_{t_2}
F\left(Q^2_{\rm S}(n_{\rm ch})\right) \left( 2 + \cos2(\phi_1 - \phi_2) \right)
\eea
for small $p_t$, where in the saturation region and the semi-saturation, or BFKL kinematic region~\cite{LR} 
\bea
F\left(Q^2_{\rm S}(n_{\rm ch})\right) & = & Q^{-4}_{\rm S}(n_{\rm ch}), \nonumber  \\
F\left(Q^2_{\rm S}(n_{\rm ch})\right) & = & (\mu^4/15) Q^{-8}_{\rm S}(n_{\rm ch})
\label{Eq8}
\eea
respectively. The per final-state pair correlation quantity $f_{2(2I\!\!P)}(Q^2_{\rm S};1,2)$ defined in Eqs.~(\ref{Eq7}) and (\ref{Eq8}) was taken directly from the LR-model~\cite{LR}. In the first line of Eq.~(\ref{Eq8}) the saturation limit estimate of the $Q_{\rm T}$ momentum integral was assumed while in the second line this integral was calculated in the BFKL kinematic region. The saturation scale is determined by the total multiplicity in the collision, $n_{\rm ch}$~\cite{KN}, and $\mu^2$ is experimentally determined to be between 0.8~GeV$^2$ and 1.6~GeV$^2$~\cite{LR}. Factor $(n_{\rm s}^{\prime} - 1)/n_{\rm s}^{\prime}$ imposes pair-number normalization by requiring $\Delta\rho \rightarrow 0$ if $\rho_{2(2I\!\!P)}(n_{\rm s}^{\prime};1,2)$ can be expressed as a product of single-gluon densities, where the integrated two-gluon density in Eq.~(\ref{Eq5}) equals $n_{\rm s}^{\prime}(n_{\rm s}^{\prime} - 1)$ per collision~\cite{neutrals}.

For p+p collisions with an arbitrary, fixed number ($m \geq 1$) of parton showers and soft-component multiplicity $n_{\rm s}$, the two-gluon density is
\bea
\rho_{m(mI\!\!P)}(n_{\rm s};1,2) & = & \frac{n_{\rm s} - 1}{n_{\rm s}}
\rho_{m(mI\!\!P)}(n_{\rm s};1) \rho_{m(mI\!\!P)}(n_{\rm s};2) \nonumber \\
 & + & \Delta\rho_{m(mI\!\!P)}(n_{\rm s};1,2).
\label{Eq9}
\eea
Assuming the correlated gluon pair density scaling in Ref.~\cite{LR}, the preceding correlated gluon density can be approximated as
\bea
& & \hspace{-0.3in} \Delta\rho_{m(mI\!\!P)}(n_{\rm s};1,2)  \approx  \frac{1}{2} m(m-1)
\Delta\rho_{2(mI\!\!P)}(n_{\rm s};1,2)
\label{Eq10}
\eea
where
\bea
& & \hspace{-0.3in} \Delta\rho_{2(mI\!\!P)}(n_{\rm s};1,2)  \approx 
\Delta\rho_{2(2I\!\!P)}(n_{\rm s}^{\prime};1,2) \nonumber \\
 &  & \hspace{-0.3in} = \frac{n_{\rm s}^{\prime} - 1}{n_{\rm s}^{\prime}}
\rho_{2(2I\!\!P)}(n_{\rm s}^{\prime};1) \rho_{2(2I\!\!P)}(n_{\rm s}^{\prime};2)
f_{2(2I\!\!P)}(Q^2_{\rm S};1,2)
\label{Eq11}
\eea
using Eqs.~(\ref{Eq5}) and (\ref{Eq6}), and $n_{\rm s}^{\prime} = 2 n_{\rm s} / m$ is the soft-component multiplicity from any two of the $m$ parton showers~\cite{fluctuations}. If the ensemble of p+p collisions with soft-component multiplicity $n_{\rm s}$ includes collisions with varying numbers of parton showers, $m = 1,2,\cdots M$, then the two-gluon density for this collision event ensemble is given by the sum
\bea
\rho(n_{\rm s};1,2) & = & \sum_{m=1}^M \tilde{P}_m(n_{\rm ch}) 
\rho_{m(mI\!\!P)}(n_{\rm s};1,2)
\label{Eq12}
\eea
where $\tilde{P}_m(n_{\rm ch})$ is the probability that a collision producing multiplicity $n_{\rm ch}$ has $m$ parton showers (see Fig.~\ref{Fig1}b).

The reference density used in data analysis is obtained by histogramming charged-particle pairs from mixed events within a fixed multiplicity event ensemble where collisions with different numbers of parton showers may be mixed. The corresponding two-gluon reference density in the parton shower model is given by the uncorrelated product
\bea
\rho_{\rm ref}(n_{\rm s};1,2) & = & \frac{n_{\rm s} - 1}{n_{\rm s}}
\sum_{m=1}^M \tilde{P}_m(n_{\rm ch}) \rho_{m(mI\!\!P)}(n_{\rm s};1)
\nonumber \\
& & \hspace{-0.3in} \times \sum_{m^{\prime}=1}^M \tilde{P}_{m^{\prime}}(n_{\rm ch}) \rho_{m^{\prime}(m^{\prime}I\!\!P)}(n_{\rm s};2)
\label{Eq13}
\eea
assuming a large number of collisions.

The normalized, per final-state pair correlation corresponding to that constructed in the analysis of data is given by
\bea
 & & \hspace{-0.3in} \left[ \Delta\rho / \rho_{\rm ref} \right] (n_{\rm s};1,2)
 = 
\frac{\rho(n_{\rm s};1,2) - {\cal N} \rho_{\rm ref}(n_{\rm s};1,2)}
{{\cal N} \rho_{\rm ref}(n_{\rm s};1,2)}
\label{Eq14}
\eea
where normalization factor ${\cal N}$ ensures that $\Delta\rho / \rho_{\rm ref} \rightarrow 0$ when correlation function $f_{2(2I\!\!P)}$ in Eq.~(\ref{Eq7}) vanishes. Using Eqs.~(\ref{Eq12}) and (\ref{Eq13}), substituting the required expression for ${\cal N}$, and using the single-gluon density definition in Eq.~(\ref{Eq4}), the normalized correlation quantity $\Delta\rho / \rho_{\rm ref}$ is given by
\begin{widetext}
\bea
\left[ \Delta\rho / \rho_{\rm ref} \right] (n_{\rm s};1,2)
 & = &
\frac{
\sum_{m=1}^M \tilde{P}_m(n_{\rm ch}) \frac{1}{2} m(m-1) \left( 
\frac{n_{\rm s}^{\prime} - 1}{n_{\rm s}^{\prime}} \right) 
\rho_{2(2I\!\!P)}(n_{\rm s}^{\prime};1) \rho_{2(2I\!\!P)}(n_{\rm s}^{\prime};2)
f_{2(2I\!\!P)}(Q^2_{\rm S};1,2) 
}{
  \sum_{m=1}^M \tilde{P}_m(n_{\rm ch}) \frac{n_{\rm s} - 1}{n_{\rm s}}
\rho_{m(mI\!\!P)}(n_{\rm s};1) \rho_{m(mI\!\!P)}(n_{\rm s};2)
}.
\label{Eq15}
\eea
\end{widetext}
In order to compare Eq.~(\ref{Eq15}) with the $p_t$-integral p+p angular correlation data at 200~GeV~\cite{TomDuncan} and 7~TeV~\cite{CMSpp7TeV} both numerator and denominator must be integrated over transverse momenta $p_{t_1}$ and $p_{t_2}$. Using the approximations in Eqs.~(\ref{Eq4}) and (\ref{Eq7}) the integrals result in the mean $p_t^2$
\bea
\langle p^2_{t_1} \rangle_{n_{\rm ch}} & = & \int_{0}^{\infty} dp_{t_1} p^2_{t_1} 
\hat{\rho}_{\rm s}(1)
/ \int_{0}^{\infty} dp_{t_1} \hat{\rho}_{\rm s}(1)
\label{Eq16}
\eea
and similarly for $\langle p^2_{t_2} \rangle_{n_{\rm ch}}$. The $n_{\rm ch}$-dependence of the shape of the soft-component of the 200~GeV single-particle density at mid-rapidity, estimated in Ref.~\cite{Tom-pp}, was included resulting in an 18\% decrease in $\langle p^2_t \rangle$ from the lowest to the highest multiplicities considered here. For the 7~TeV data the minimum-bias $p_t$ distribution data from CMS~\cite{CMSdNdeta} were used to estimate the soft-component $\langle p^2_{t} \rangle$. With this $p_t$ integration Eq.~(\ref{Eq15}) simplifies to
\begin{widetext}
\bea
\left[ \Delta\rho / \rho_{\rm ref} \right] (n_{\rm s};1,2)
& = &
\sum_{m=1}^M \tilde{P}_m(n_{\rm ch}) (m-1)
\left( \frac{2n_{\rm s}/m - 1}{2(n_{\rm s} - 1)} \right)
\langle p^2_t \rangle^2_{n_{\rm ch}} F\left(Q^2_{\rm S}(n_{\rm ch})\right)
\left( 2 + \cos2(\phi_1 - \phi_2) \right).
\label{Eq17}
\eea

In the correlation analysis of data presented in Refs.~\cite{TomDuncan,axialCI} a per final-state particle normalization was used where the ratio $\Delta\rho / \rho_{\rm ref}$ was multiplied by the efficiency corrected, single-particle density $n_{\rm ch}/(2\pi\, \Delta\eta)$. The resulting correlation data were described with a fitting model where the quadrupole correlation model element was defined as $2A_{\rm Q} \cos2(\phi_1 - \phi_2)$. The corresponding quadrupole correlation amplitude computed here using Eq.~(\ref{Eq17}) is therefore given by
\bea
A_{\rm Q}(n_{\rm ch}) & = & \frac{1}{2} \frac{n_{\rm ch}}{2\pi ~ \Delta\eta}
\left[ \sum_{m=1}^M \tilde{P}_m(n_{\rm ch}) (m-1) 
\left( \frac{2n_{\rm s}/m - 1}{2(n_{\rm s} - 1)} \right) \right]
\langle p^2_t \rangle^2_{n_{\rm ch}} F\left(Q^2_{\rm S}(n_{\rm ch})\right)
\label{Eq18}
\eea
\end{widetext}
for $n_{\rm ch} \geq 2$ and $n_{\rm s} > 1$.

The minimum-bias average quadrupole amplitude for the data~\cite{TomDuncan} was obtained from the total-pair weighted average of the minimum-bias event ensemble, where
\bea
A_{\rm Q,mb} & = & \sum_{n_{\rm ch} \geq 2} W_{n_{\rm ch}}
\frac{\langle n_{\rm ch} \rangle / \Delta\eta}{n_{\rm ch} / \Delta\eta}
A_{\rm Q}(n_{\rm ch}), \nonumber \\
W_{n_{\rm ch}} & = &
\frac{
N_{\rm events}(n_{\rm ch}) n_{\rm ch} (n_{\rm ch} - 1)
}{
\sum_{n_{\rm ch}^{\prime} \geq 2} N_{\rm events}(n_{\rm ch}^{\prime})
n_{\rm ch}^{\prime} (n_{\rm ch}^{\prime} - 1)
},
\label{Eq19}
\eea
and $N_{\rm events}(n_{\rm ch})$ is the number of p+p collision events with multiplicity $n_{\rm ch}$ in acceptance $\Delta\eta$ in the minimum-bias sample of collisions. For the latter the $|\eta| < 0.5$ UA5~\cite{UA5} and CMS~\cite{CMSFreq} distributions were assumed. The same minimum-bias averaging procedure was used for the LR-model calculations using Eq.~(\ref{Eq18}) and the experimental frequency distributions.

Pomeron probabilities were estimated in Sec.~\ref{SecII}. Calculation of Eq.~(\ref{Eq18}) requires saturation scale estimates which are discussed next.

\section{Expected saturation scales}
\label{SecIV}

Theoretical discussions of gluon saturation in hadrons and nuclei are plentiful in the literature, for example Refs.~\cite{RajuDusling,KN,Dumitru,LR2}. Simply stated, when the gluon density projection onto the 2D plane transverse to the beam direction exceeds the inverse area per gluon, $[\alpha_{\rm s}(Q^2)\pi/Q^2]^{-1}$, where $\alpha_{\rm s}(Q^2)$ is the running coupling constant of the strong interaction~\cite{alphas}, the initial-state gluons start to overlap and their density begins to saturate due to gluon fusion ($g+g \rightarrow g$). The onset of saturation occurs at transverse momentum scale $Q^2_{\rm S}$. For nucleus + nucleus (A+A) collisions the number of gluons with kinematic quantities $(x,Q^2)$ in terms of the gluon structure function for protons and the number of nucleon participants in the A+A collision is approximately proportional to $xG(x,Q^2) N_{\rm part}$, where $xG(x,Q^2) \approx \gamma \ln (Q^2/\Lambda_{\rm QCD}^2)$. At $x \approx 10^{-2}$ (RHIC) and $3 \times 10^{-4}$ (LHC) $\gamma \approx 0.88$ and 2.3, respectively~\cite{HERA}, where $\Lambda_{\rm QCD}$ is assumed to be 0.2~GeV~\cite{KN}. Saturation occurs when
\bea
\frac{\pi R_{\rm A}^2}{\alpha_{\rm s}(Q^2_{\rm S})\pi/Q^2_{\rm S}}
& = &
\gamma C \ln \left( \frac{Q^2_{\rm S}}{\Lambda_{\rm QCD}^2} \right) N_{\rm part},
\label{Eq20}
\eea
where $\pi R_{\rm A}^2$ is the nucleus + nucleus transverse overlap area and $C$ is a proportionality constant. For most-central (head-on) Au+Au collisions at $\sqrt{s_{\rm NN}} = 200$~GeV $Q^2_{\rm S}$ was estimated in Refs.~\cite{KN,Dumitru} to be in the range from 1 to 2~GeV$^2$ which can be used to determine coefficient $C$.

The nucleon participant areal density in Eq.~(\ref{Eq20}) for Au+Au collisions is proportional to $N_{\rm part}^{1/3}$ where~\cite{MCG}
\bea
\frac{N_{\rm part}}{\pi R_{\rm A}^2} & \approx & 0.37({\rm fm}^{-2})
N_{\rm part}^{1/3} ,
\label{Eq21}
\eea
which in the minimum-bias p+p collision limit ($N_{\rm part} = 2$) equals $0.47~{\rm fm}^{-2}$. The latter is approximately equal to $N_{\rm part}/\sigma_{\rm NN}$ = 2/42~mb = 0.48~fm$^{-2}$ where $\sigma_{\rm NN}$ = 42~mb is the p+p total inelastic cross section at $\sqrt{s}$ = 200~GeV~\cite{PDG}. Iterative solution of Eq.~(\ref{Eq20}) for $Q^2_{\rm S}$ in the p+p limit, assuming the above range of values for coefficient $C$, results in $Q^2_{\rm S}$ for 200~GeV minimum-bias p+p collisions ranging from 0.204~GeV$^2$ to 0.387~GeV$^2$. At 7~TeV with $\sigma_{\rm NN} \approx 90$~mb~\cite{PDG} solution of Eq.~(\ref{Eq20}) yields a range of $Q^2_{\rm S}$ from 0.244~GeV$^2$ to 0.468~GeV$^2$.

\begin{figure}[t]
\includegraphics[keepaspectratio,width=3.5in]{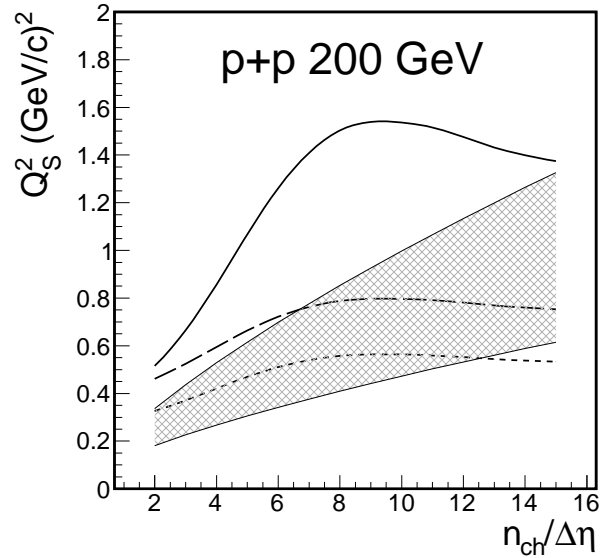}
\caption{\label{Fig3}
Range of saturation scales $Q^2_{\rm S}(n_{\rm ch})$ from theoretical predictions as discussed in Sec.~\ref{SecIV} (shaded band). Saturation scales $Q^2_{\rm S}(n_{\rm ch})$ from the Levin and Rezaeian model application in Eq.~(\ref{Eq18}) which are required to describe the experimental 200 GeV p+p quadrupole correlation in Eq.~(\ref{Eq23})~\cite{TomDuncan} assuming three saturation limits [Eq.~(\ref{Eq8})]. Solid, long-dashed and short-dashed curves correspond to the full saturation limit and the semi-saturation limit (BFKL region) with $\mu^2$ = 1.6~GeV$^2$ and 0.8~GeV$^2$, respectively.}
\end{figure}

The saturation scale for multiplicity dependent p+p collisions was estimated by: (1) assuming the soft-component of the final-state charged particle multiplicity provides a reasonable proxy for the relevant number of low-$x$ gluons in the initial state (e.g. LPHD)~\cite{LPHD}, and (2) assuming a fixed transverse area for p+p collisions (e.g. $\sigma_{\rm NN}$). Using Eq.~(\ref{Eq2}) the saturation condition for multiplicity dependent p+p collisions can then be written as
\bea
\frac{n_{\rm gluon}}{\Delta\eta} & \approx & \frac{3}{2}
\frac{n_{\rm s}}{\Delta\eta} \nonumber \\
& = & \frac{\frac{3}{2} n_{\rm ch}/\Delta\eta}
{1 + \alpha n_{\rm ch}/(\epsilon \Delta\eta)}
=
C^{\prime} \frac{\sigma_{\rm NN}}{\alpha_{\rm s}(Q^2_{\rm S})\pi/Q^2_{\rm S}},
\label{Eq22}
\eea
where factor 3/2 accounts for neutral particle production and $C^{\prime}$ is another proportionality constant which is determined by the saturation scale for the average minimum-bias p+p collision with $n_{\rm ch}/\Delta\eta = \langle n_{\rm ch} \rangle / \Delta\eta =2.5$ at 200~GeV~\cite{UA5} and 5.78 at 7~TeV~\cite{CMSFreq,CMSdNdeta}. Saturation scale $Q^2_{\rm S}(n_{\rm ch})$ was obtained by iterative solution of Eq.~(\ref{Eq22}). The resulting range of theoretically expected $Q^2_{\rm S}$ as functions of p+p multiplicity for 200~GeV collisions is shown by the monotonically increasing shaded band in Fig.~\ref{Fig3}.

\section{Results at 200 GeV}
\label{SecV}

The multiplicity dependent trend of the preliminary $A_{\rm Q,exp}(n_{\rm ch})$ data~\cite{TomDuncan} for 200~GeV p+p collisions with $n_{\rm ch} \in [2,15]$ and $|\eta| < 0.5$ can be approximated with
\bea
A_{\rm Q,exp}(n_{\rm ch}) & \approx & \frac{n_{\rm s}}{n_{\rm ch}}
\left[ a_0 + a_1 \frac{n_{\rm s}}{\Delta\eta} + a_2
\left( \frac{n_{\rm s}}{\Delta\eta} \right)^2  \right],
\label{Eq23}
\eea
where $a_0 = -0.000267$, $a_1$ = 0.00048, $a_2$ = 0.0000243, and the functional relation between $n_{\rm s}$ and $n_{\rm ch}$ is given in Eq.~(\ref{Eq2}). The minimum-bias weighted average quadrupole amplitude using Eq.~(\ref{Eq19}) and the $|\eta| \leq 0.5$ UA5 multiplicity frequency distribution data~\cite{UA5} is $A_{\rm Q,mb} = 0.00135 \pm 0.00009$.

\begin{figure}[t]
\includegraphics[keepaspectratio,width=3.7in]{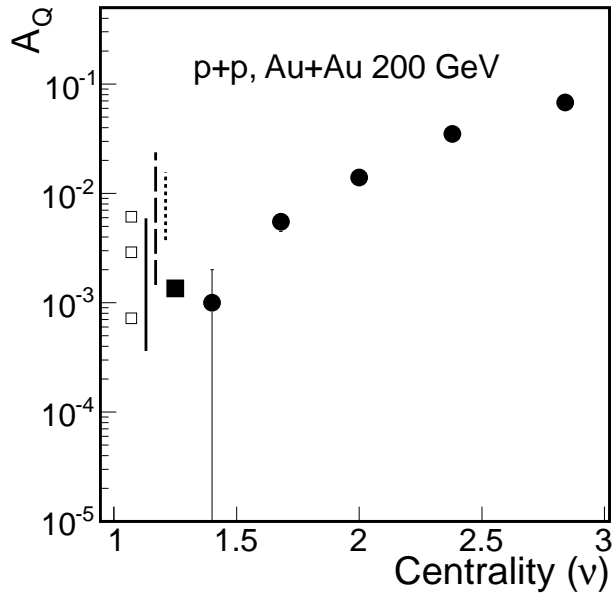}
\caption{\label{Fig4}
Quadrupole amplitudes for 200~GeV minimum-bias p+p and Au+Au centrality dependent collisions (solid circles)~\cite{axialCI}. The p+p datum~\cite{TomDuncan} is shown by the solid square symbol. Error bars, if not sown, are smaller than the symbols. LR-model results assuming fixed $Q_{\rm S}^2 = 0.6$~GeV$^2$ (see text) are shown by the open squares. The LR-model results for minimum-bias p+p collisions assuming the theoretical saturation scale range in Fig.~\ref{Fig3} for the semi-saturated $\mu^2$ = 0.8~GeV$^2$, 1.6~GeV$^2$, and fully saturated limits are shown by the solid, long-dashed and short-dashed vertical lines, respectively. The LR-model results are offset from $\nu$ = 1.25 for clarity.}
\end{figure}

Prior to the measurement of the p+p quadrupole amplitude~\cite{TomDuncan} the only available estimates of its value were from extrapolations to the single nucleon + nucleon (N+N) limit of quadrupole amplitudes measured in Au+Au collisions at 200~GeV per colliding N+N pair~\cite{axialCI}. The 200~GeV Au+Au quadrupole amplitudes including statistical and systematic uncertainties for the five most-peripheral collision centrality bins are shown in Fig.~\ref{Fig4} (solid circles) as a function of centrality measure $\nu$~\cite{axialCI}. In nucleus + nucleus collisions multiplicity at mid-rapidity is used to estimate the degree of overlap, or {\em centrality}, of the colliding nuclei. Parameter $\nu$ is estimated via Monte Carlo methods~\cite{MCG} as the average number of N+N collisions per incoming nucleon in the beam nucleus which participates in the interaction. For p+p collisions the nominal value $\nu = 1$ extends to an effective value $\nu = 1.25$ due to asymmetry in the minimum-bias multiplicity frequency distribution~\cite{MCG,TomMCG}. Extrapolation of the Au+Au quadrupole amplitudes to the p+p limit suggests a range of values from 0.0 to 0.002. The measured quadrupole amplitude is shown in Fig.~\ref{Fig4} by the solid square symbol which is within reasonable extrapolation limits from the Au+Au data.

The LR-model prediction for the minimum-bias average quadrupole amplitude was calculated using Eqs.~(\ref{Eq8}), (\ref{Eq18}) and (\ref{Eq19}), the Pomeron probabilities shown in Fig.~\ref{Fig1}b, the above UA5 measured frequency distribution, and parameters $\langle n_{\rm ch} \rangle / \Delta\eta$ = 2.5 ($\pm$5\% uncertainty~\cite{UA5}), $\langle p_t^2 \rangle \in [0.164,0.201]$~(GeV/$c$)$^2$ decreasing with $n_{\rm ch}$~\cite{Tom-pp}, and $Q^2_{\rm S}$ = 0.6~GeV$^2$ (from Ref.~\cite{LR} which assumed LHC energies). The resulting minimum-bias average quadrupole amplitude in the LR-model depends on the degree of gluon saturation assumed for the $Q_{\rm T}$ momentum integral which is included in the quantity $F(Q^2_{\rm S})$ in Eq.~(\ref{Eq8}). Assuming a constant value for $Q^2_{\rm S} = 0.6$~GeV$^2$ results in $A_{\rm Q,mb}$ varying from 0.00072 to 0.0029 for $\mu^2$ = 0.8~GeV$^2$ and 1.6~GeV$^2$ respectively, in the semi-saturation (BFKL) limit, and increasing to 0.0061 in the full saturation limit. These three predicted values are shown in Fig.~\ref{Fig4} by the open square symbols. The measured amplitude 0.00135 is intermediate between the two predictions which assumed semi-saturation of the gluon distribution. The full saturation prediction is a factor of 4.5 larger than the data.

LR-model predictions can also be derived assuming the range of theoretically expected saturation scales shown in Fig.~\ref{Fig3}. Using Eqs.~(\ref{Eq8}), (\ref{Eq18}) and (\ref{Eq19}) with the semi-saturated and fully saturated limits, and assuming the upper and lower limits of the theoretically expected $Q_{\rm S}^2(n_{\rm ch})$ results in the three ranges for the minimum-bias p+p quadrupole amplitude shown on the left side of Fig.~\ref{Fig4}. The ranges of $A_{\rm Q}$ for the semi-saturated limit with $\mu^2$ = 0.8~GeV$^2$ and 1.6~GeV$^2$ are shown by the solid and dashed vertical lines, respectively, which encompass the data point. The range for the fully saturated limit is indicated by the dotted line. The factor of two uncertainty in $\mu^2$ results in a factor of four offset between the solid and dashed lines. The approximate factor of two uncertainty in $Q_{\rm S}^2$ estimated in Sec.~\ref{SecIV} results in an approximate factor of 4 (16) uncertainty for the saturation (semi-saturation) limit estimate. Uncertainties in the LR-model predictions (e.g. due to Pomeron exchange probabilities, $Q_{\rm T}$ integration, saturation scales) are multiplicative not additive, which motivates the $\log$ scale in Fig.~\ref{Fig4}. The $\log$-scale is intended to emphasize the fact that the uncertainty range for the LR-model might not encompass the data and therefore the model can, in principle, be falsified.

It is also interesting to compare the theoretically expected saturation scales with the required $Q^2_{\rm S}(n_{\rm ch})$ which, in the LR-model Eq.~(\ref{Eq18}), reproduces the experimental quadrupole correlation in Eq.~(\ref{Eq23}). Solving Eq.~(\ref{Eq18}) for $Q^2_{\rm S}(n_{\rm ch})$ in both the full saturation and semi-saturation limits expressed in Eq.~(\ref{Eq8}), and using the measured quadrupole correlation amplitude represented in Eq.~(\ref{Eq23}), results in the three curves for $Q^2_{\rm S}(n_{\rm ch})$ shown in Fig.~\ref{Fig3}. In this figure the required $Q^2_{\rm S}(n_{\rm ch})$ values in the full saturation limit and in the semi-saturation limit with $\mu^2$ = 1.6~GeV$^2$ and 0.8~GeV$^2$ are shown by the solid, long-dashed and short-dashed curves, respectively, as functions of final-state charged particle multiplicity per unit $\eta$. Within the LR-model, successful description of the quadrupole correlation data requires a generally increasing saturation scale with increasing $n_{\rm ch}$. The overlap between the long and short-dashed lines and the theoretical band means that the LR-model in the semi-saturation limit is capable of describing $A_{\rm Q,exp}(n_{\rm ch})$ using saturation scales which lie within the theoretically expected range.

It is worth noting that the quadrupole correlation measurements~\cite{TomDuncan}, the LR-model calculations, and the theoretical estimates of the gluon saturation scale are all independent. The overlap of these three results shown in Figs.~\ref{Fig3} and \ref{Fig4} is interesting given the model uncertainties in the momentum integrals in the saturation region. In Fig.~\ref{Fig3} both the magnitudes and the generally increasing trends of $Q^2_{\rm S}$ with multiplicity obtained in this application of the LR-model in the BFKL region (semi-saturation limit) are shown to be similar to theoretical expectations.

\section{Results at 7 TeV}
\label{SecVI}

\begin{figure}[t]
\includegraphics[keepaspectratio,width=3.7in]{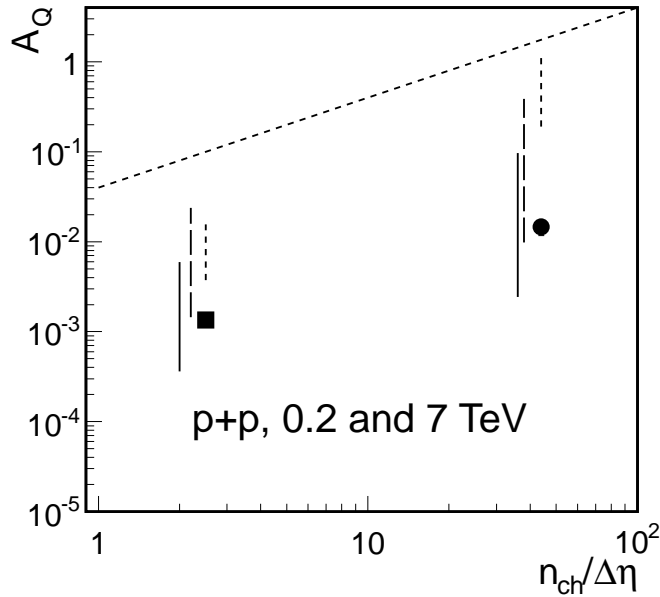}
\caption{\label{Fig5}
Measured and predicted quadrupole amplitudes for 200~GeV minimum-bias p+p collisions repeated from Fig.~\ref{Fig4} (solid square and vertical lines near $n_{\rm ch}/\Delta\eta= 2.5$) in comparison with similar predictions and measured quadrupole amplitude (solid circle symbol) for the $N^{\rm offline}_{\rm trk} \ge 110$, $p_t > 0.1$~GeV/$c$ 7~TeV p+p collision data from CMS~\cite{CMSpp7TeV}. LR-model results assuming the semi-saturated $\mu^2$ = 0.8~GeV$^2$, 1.6~GeV$^2$, and fully saturated limit are shown by the solid, long-dashed and short-dashed vertical lines near $n_{\rm ch}/\Delta\eta= 44$, respectively. The LR-model results are horizontally offset for clarity. The upper dashed line indicates the maximum allowed quadrupole amplitude discussed in the text.}
\end{figure}

The long-range correlations on relative pseudorapidity reported by the CMS collaboration for 7~TeV p+p collisions are most apparent for the intermediate momentum charged particles from higher multiplicity collisions (see Fig. 7 in Ref.~\cite{CMSpp7TeV}). The analysis of the 2D ($\eta,\phi$) correlations in Ref.~\cite{LRQuadpp7TeV} found a significant quadrupole correlation for the $N^{\rm offline}_{\rm trk} \ge 110$, $p_t > 0.1$~GeV/$c$ and $1.0 \le p_t < 3.0$~GeV/$c$ data, but not for the minimum-bias collisions. Particle number $N^{\rm offline}_{\rm trk}$ in \cite{CMSpp7TeV} refers to offline reconstructed tracks within $|\eta| < 2.4$ and with $p_t > 0.4$~GeV/$c$ and is not corrected for detector and tracking algorithm inefficiencies. The LR-model density in Eq.~(22) of Ref.~\cite{LR} is only applicable for smaller values of $p_t$; therefore the LR-model prediction for the quadrupole amplitude in Eq.~(\ref{Eq18}) may only be applied to the $p_t > 0.1$~GeV/$c$ correlation data from CMS.

In Ref.~\cite{LRQuadpp7TeV} a quadrupole correlation of the form $A_{\rm Q} \cos 2(\phi_1 - \phi_2)$ was found to describe the CMS correlation quantity $R(\eta_1 - \eta_2,\phi_1 - \phi_2)$ which includes prefactor $(\langle N^{\rm offline}_{\rm trk} \rangle - 1) = 116.8$ for the $N^{\rm offline}_{\rm trk} \ge 110$ collisions. In order to convert the quadrupole amplitude in Ref.~\cite{LRQuadpp7TeV} to the definition used here a factor of 2 [for present definition 2$A_{\rm Q} \cos 2(\phi_1 - \phi_2)$] and the preceding prefactor used by CMS must be divided out, and a prefactor, $n_{\rm ch}/(2\pi\, \Delta\eta)$, for the higher multiplicities must be applied. The latter was estimated by applying a $p_t$-acceptance correction (because of the $p_t > 0.4$~GeV/$c$ limit), estimated from minimum-bias collisions~\cite{CMSdNdeta} as $(\langle n_{\rm ch} \rangle / \Delta\eta)/(\langle N^{\rm corrected}_{\rm trk} \rangle / \Delta\eta)$ = $(5.78 \pm 0.23)/(17.8/4.8)$ = $1.56 \pm 0.06$, to the average, corrected higher multiplicity $\langle N^{\rm corrected}_{\rm trk} \rangle / \Delta\eta = 136.1/4.8$~\cite{CMSpp7TeV}. The resulting $n_{\rm ch}/(2\pi\, \Delta\eta)$ equals $44.2/2\pi = 7.03$. The resulting quadrupole amplitude from \cite{LRQuadpp7TeV} for the high multiplicity $p_t > 0.1$~GeV/$c$ data is $0.0146 \pm 0.0007$(stat) $\pm 0.0022$(syst) and is shown in Fig.~\ref{Fig5} by the solid circle symbol. 

LR-model quadrupole correlation predictions for these data were computed for the average, corrected multiplicity estimated in the preceding paragraph, $n_{\rm ch}/\Delta\eta = 44$, corresponding to the $N^{\rm offline}_{\rm trk} \ge 110$ data. Using the Pomeron exchange probabilities obtained in Sec.~\ref{SecII} for each assumed value of hard-scattering fraction $\alpha$, results in probability factors [quantity in square brackets in Eq.~(\ref{Eq18})] ranging between 0.624 and 0.720.  The mean-$p_t^2$ equals 0.498~(GeV/$c$)$^2$ from the minimum-bias $p_t$ spectrum reported in \cite{CMSdNdeta}. Iterative solutions of Eq.~(\ref{Eq22}) for $n_{\rm ch}/\Delta\eta = 44$ and for the three assumed values of $\alpha = 0.0105$, 0.02 and 0.03 result in expected ranges of $Q^2_{\rm S} \in [0.84,1.82]$~GeV$^2$, [0.76,1.64]~GeV$^2$, and [0.70,1.50]~GeV$^2$, respectively. The resulting ranges of the predicted quadrupole amplitudes for the semi-saturated limit with $\mu^2$ = 0.8~GeV$^2$ and 1.6~GeV$^2$, and for the full saturation limit are shown respectively by the vertical solid, long-dashed and short-dashed lines in Fig.~\ref{Fig5} for $n_{\rm ch}/\Delta\eta = 44$. The semi-saturated, BFKL kinematic region predictions encompass the data point. The corresponding 200~GeV results from Fig.~\ref{Fig4} are repeated in Fig.~\ref{Fig5} to facilitate direct comparison.  The uncertainties are relatively larger for the 7~TeV predictions than for the 200~GeV predictions because of the additional uncertainty in the hard-scattering parameter $\alpha$. The upper dashed line indicates the maximum allowed quadrupole amplitude explained in the next section.

\section{Discussion}
\label{SecVII}

From the results in Figs.~\ref{Fig4} and \ref{Fig5} it can be seen that even with the relatively large uncertainties in the model predictions, inclusion of the measured value within the predicted range is not guaranteed. For example, if the Pomeron exchange probabilities, the mean saturation scale, and/or the momentum integration over BFKL Pomerons in the saturation region (quantity $\langle q_{\rm T}^{-4} \rangle$ in \cite{LR}) were very different from that estimated here and in Ref.~\cite{LR}, the uncertainty bands in Figs.~\ref{Fig4} and \ref{Fig5} would be shifted up or down, while remaining within the overall allowed range discussed below, and might not encompass the measurements. The reduced uncertainties in the Pomeron exchange probabilities and saturation scale obtained in this analysis enable the present application of the LR-model to be falsified in principle. The results in Figs.~\ref{Fig4} and \ref{Fig5} show however that the LR-model for the BFKL kinematic region (solid and long-dashed lines) assuming the momentum integrals estimated in Ref.~\cite{LR} and the $\tilde{P}_m(n_{\rm ch})$ and $Q_{\rm S}^2(n_{\rm ch})$ estimated here is not excluded by the p+p quadrupole data at either energy. Furthermore, from Fig.~\ref{Fig5} it is seen that the LR-model application in the BFKL region accurately predicts the relative increase in the amplitudes of the quadrupole correlation for these two cases. The latter result is interesting given that several factors in Eq.~(\ref{Eq18}) vary an order-of-magnitude or more between the two cases and that the relevant numbers of Pomeron exchanges differ significantly. It should also be pointed out that with improved calculations of momentum integrals $\langle q_{\rm T}^{-4} \rangle$ and $\langle \langle Q_{\rm T}^4 \rangle \rangle $~\cite{LR} in the saturation region, future applications of the LR-model might be excluded by the data.

The single-gluon distribution of the LR-model includes a $\cos 2\phi$ dependence relative to the event-wise momentum transfer direction $\hat{Q}_{\rm T}$~\cite{LR}. It is therefore appropriate to consider the single-particle azimuthal anisotropy amplitude $v_2$ from this model, where $dn_{\rm ch}/d\phi \propto [1+2v_2 \cos 2 (\phi - \hat{Q}_T)]$. For an ensemble of collisions having this single-particle distribution, the two-gluon azimuthal correlation results from averaging over the random momentum transfer directions and is proportional to $2v_2^2 \cos 2(\phi_1 - \phi_2)$. Including the per-particle normalization factor [see Eq.~(\ref{Eq18})], the quadrupole amplitude $A_{\rm Q}$ and single-particle anisotropy amplitude $v_2$ are related by
\bea
A_{\rm Q} & = & \frac{n_{\rm ch}}{2\pi \Delta\eta} v_2^2,
\label{Eq24}
\eea
assuming no other mechanisms contribute to the quadrupole. In the semi-saturation limit with fixed $Q^2_{\rm S}$ = 0.6~GeV$^2$ the $v_2$ azimuthal anisotropy amplitudes for the 200~GeV p+p minimum-bias average are 0.043 and 0.085 corresponding to $\mu^2$ = 0.8~GeV$^2$ and 1.6~GeV$^2$, respectively. This range of $v_2$ values is comparable to $p_t$-integral $v_2$ amplitudes for nucleus + nucleus collisions~\cite{AAv2}. Assuming the quadrupole correlation in the 200~GeV p+p data is similarly generated by event-wise anisotropy in the single-particle distribution, the $p_t$-integral minimum-bias average $v_2$ = 0.058. The p+p multiplicity dependent quadrupole correlation~\cite{TomDuncan} results in $v_{\rm 2,exp}(n_{\rm ch})$ varying from approximately 0.05 to 0.07 for multiplicities from 2 to 15 per unit $\eta$, respectively. Similarly, for the 7~TeV high multiplicity correlation data $v_{\rm 2,exp}$ equals 0.046~\cite{LRQuadpp7TeV}.

From the above definition of $v_2$, $|v_2| \leq 0.5$ is required for non-negative probabilities and therefore $0 \leq A_{\rm Q} \leq (dn_{\rm ch}/2\pi\Delta\eta)(0.5)^2 \approx 0.1$ for minimum-bias 200~GeV p+p collisions and $\approx 1.8$ for the high multiplicity 7~TeV p+p collisions. In the absence of numerical estimates of the Pomeron exchange probabilities $\tilde{P}_m$, momentum integrals $F(Q_{\rm S}^2)$, and saturation scale the quadrupole amplitude in the LR-model can only be constrained within the range [0.0,0.1] and [0.0,1.8] for the two cases studied here. The upper limit for the quadrupole amplitude as a function of multiplicity is shown by the upper dashed line in Fig.~\ref{Fig5}.

Although the quadrupole correlation studied here is small relative to other features of 2D p+p angular correlations, e.g. jets and dijets~\cite{LRQuadpp7TeV,TomDuncan,axialCI}, the single-particle anisotropy amplitude $v_2$ is very similar to that reported for high-energy heavy-ion collisions~\cite{AAv2}. It is worth noting that the quadrupole correlation in p+p collisions is not necessarily restricted to high-multiplicity collisions but that it is also statistically significant at low multiplicities as reported for the 200~GeV p+p data~\cite{TomDuncan}. The quadrupole correlation therefore reflects significant, non-negligible aspects of high-energy p+p collision dynamics.

\section{Conclusions}
\label{SecVIII}

Long-range correlations on relative pseudorapidity and at small relative azimuth appear to be a ubiquitous feature in the hadron production from high energy p+p, p+A and A+A collisions. Phenomenological analysis of correlation data showed that for p+p and p+A collisions this long-range $\eta$ structure together with a portion of the away-side ($|\phi_1 - \phi_2| \approx \pi$) ridge~\cite{ALICEpPb,PHENIXdAu} can be accurately described with an $\eta$-independent azimuthal quadrupole. A number of authors have proposed gluon interference mechanisms within a pQCD framework to explain these long-range correlations. Others have proposed that these correlation structures result from collective flow generated by multiple, secondary collisions between the partons produced in the early stage of the hadronic collision.

In the present work the pQCD, Pomeron exchange model of Levin and Rezaeian~\cite{LR} using parameters estimated here was compared to recent 200~GeV p+p minimum-bias collision data from the STAR Collaboration and to the high multiplicity 7~TeV p+p collision data from the CMS Collaboration. It was shown that the uncertainty ranges of predicted quadrupole amplitudes, particularly those assuming the semi-saturated BFKL kinematic limit, encompass the experimental values at both energies. The relative increase in the measured quadrupole amplitude with increasing collision energy and multiplicity is well reproduced by the model. The reduction in model uncertainties afforded by the present estimates of Pomeron exchange probabilities and saturation scales enables the LR-model with its estimated momentum integrals in the saturation region to be falsified by these two experimental results, despite the remaining, large uncertainties in the quadrupole amplitude predictions. The present application of the LR-model was found to not be excluded by these data. These results should encourage further theoretical work on the Levin and Rezaeian BFKL Pomeron exchange, gluon interference model in the gluon saturation region.

\vspace{0.1in}

This work was supported in part by the U. S. Department of Energy, Office of Science, Office of Nuclear Physics, Office of Heavy Ion Nuclear Physics under Award Number DE-FG02-94ER40845.
 


\begin{thebibliography}{99}

\bibitem{CMSpp7TeV}
V. Khachatryan {\em et al.} (CMS Collaboration), J. High Energy Phys. 09 (2010) 091; CMS Collaboration, arXiv:1009.4122.

\bibitem{TomQuadpp7TeV}
T. A. Trainor and D. T. Kettler, arXiv:1010.3048v2 [hep-ph].

\bibitem{Bozek}
P. Bo\.zek, Eur. Phys. J. C {\bf 71}, 1530 (2011).

\bibitem{LRQuadpp7TeV}
R. L. Ray, Phys. Rev. D {\bf 84}, 034020 (2011).

\bibitem{ATLASpPb}
G. Aad {\em et al.} (ATLAS Collaboration), Phys. Rev. Lett. {\bf 110}, 182302 (2013).

\bibitem{ALICEpPb}
B. Abelev {\em et al.} (ALICE Collaboration), Phys. Lett. B {\bf 719}, 29 (2013).

\bibitem{CMSpPb}
S. Chatrchyan {\em et al.} (CMS Collaboration), Phys. Lett. B {\bf 724}, 213 (2013).

\bibitem{PHENIXdAu}
A. Adare {\em et al.} (PHENIX Collaboration), Phys. Rev. Lett. {\bf 111}, 212301 (2013).

\bibitem{TomDuncan}
D. J. Prindle (STAR Collaboration), in the Proceedings of the XLIII Int. Symp. Mult. Part. Dynamics, eds. S. Chekanov and Z. Sullivan, Illinois Institute of Technology, (IIT Press, Chicago, 2013), p.219;
T. A. Trainor and D. J. Prindle, {\em ibid.,} p.57; arXiv:1310.0408v1 [hep-ph].

\bibitem{AAv2}
B. I. Abelev {\em et al.} (STAR Collaboration), Phys. Rev. C {\bf 77}, 054901 (2008);
A. Adare {\em et al.} (PHENIX Collaboration), Phys. Rev. Lett. {\bf 105}, 062301 (2010);
B. Abelev {\em et al.} (ALICE Collaboration), Phys. Lett. B {\bf 719}, 18 (2013).

\bibitem{Boris}
B. Z. Kopeliovich, A. H. Rezaeian and I. Schmidt, Phys. Rev. D {\bf 78}, 114009 (2008).

\bibitem{LR}
E. Levin and A. Rezaeian, Phys. Rev. D {\bf 84}, 034031 (2011).

\bibitem{glasma}
T. Lappi and L. McLerran, Nucl. Phys. A {\bf 772}, 200 (2006);
F. Gelis and R. Venugopalan, Acta Phys. Polon. B {\bf 37}, 3253 (2006).

\bibitem{RajuDusling}
K. Dusling and R. Venugopalan, Phys. Rev. D {\bf 87}, 051502(R) (2013);
Phys. Rev. D {\bf 87}, 094034 (2013).

\bibitem{pPbFlow}
P. Bo\.zek and W. Broniowski, Phys. Lett. B {\bf 718}, 1557 (2013);
{\em ibid.} {\bf 720}, 250 (2013); P. Bo\.zek, Phys. Rev. C {\bf 85}, 014911 (2012).

\bibitem{LPHD}
Ya. I. Azimov, Yu. L. Dokshitzer, V. A. Khoze and S. I. Troyan, Z. Phys. C
{\bf 27}, 65 (1985); Z. Phys. C {\bf 31}, 213 (1986).

\bibitem{BFKL}
L. N. Lipatov, Sov. J. Nucl. Phys. {\bf 23}, 338 (1976);
E. A. Kuraev, L. N. Lipatov and V. S. Fadin, Sov. Phys. JETP {\bf 44}, 443 (1976);
E. A. Kuraev, L. N. Lipatov and V. S. Fadin, Sov. Phys. JETP {\bf 45}, 199 (1977);
Ya. Balitskii and L. N. Lipatov, Sov. J. Nucl. Phys. {\bf 28}, 822 (1978).

\bibitem{PomeronLoops}
E. Levin, J. Miller and A. Prygarin, Nucl. Phys. {\bf A806}, 245 (2008).

\bibitem{UA5}
R. E. Ansorge {\em et al.} (UA5 Collaboration), Z. Phys. C {\bf 43}, 357 (1989).

\bibitem{Tom-pp}
J. Adams {\em et al.} (STAR Collaboration), Phys. Rev. D {\bf 74}, 032006 (2006).

\bibitem{neutrals}
The radiated gluon yields and gluon densities are normalized to the corresponding soft-component charged hadron quantities. Neutral hadron contributions to correlations are not considered here.

\bibitem{CMSdNdeta}
V. Khachatryan {\em et al.} (CMS Collaboration), Phys. Rev. Lett. {\bf 105}, 022002 (2010).

\bibitem{KN}
D. Kharzeev and M. Nardi, Phys. Lett. B {\bf 507}, 121 (2001);
D. Kharzeev, E. Levin and M. Nardi, Nucl. Phys. A {\bf 730}, 448 (2004).

\bibitem{CMSFreq}
V. Khachatryan {\em et al.} (CMS Collaboration), J. High Energy Phys. 01 (2011) 079.

\bibitem{HERA}
M. Dittmar {\em et al.}, arXiv:0901.2504v2 [hep-ph] (2009).

\bibitem{fluctuations}
These equations are for mean values. The effect on the quadrupole correlation amplitude of event-wise statistical fluctuations in the number of gluons radiated per parton shower for collisions with fixed $n_{\rm ch}$ is 3\% or less.

\bibitem{axialCI}
G. Agakishiev {\em et al.} (STAR Collaboration), Phys. Rev. C {\bf 86}, 064902 (2012).

\bibitem{Dumitru}
A. Dumitru, F. Gelis, L. McLerran and R. Venugopalan, Nucl. Phys. A {\bf 810}, 91 (2008).

\bibitem{LR2}
E. Levin and A. Rezaeian, Phys. Rev. D {\bf 82}, 014022 (2010).

\bibitem{alphas}
S. Bethke, Prog. Part. Nucl. Phys. {\bf 58}, 351 (2007).

\bibitem{MCG}
R. L. Ray and M. S. Daugherity, J. Phys. G: Nucl. Part. Phys. {\bf 35}, 125106 (2008).

\bibitem{PDG}
Particle Data Group, J. Phys. G: Nucl. Part. Phys. {\bf 33}, 1 (2006).

\bibitem{TomMCG}
T. A. Trainor and D. J. Prindle, arXiv:hep-ph/0411217v3 (2007).

\end{thebibliography}
\end{document}